\documentclass{ws-ijmpa}

\begin{document}

\def\be{\begin{equation}}
\def\ee{\end{equation}}
\def\lp{\left(}
\def\rp{\right)}
\def\lb{\left[}
\def\rb{\right]}
\def\om{\omega}
\def\La{\Lambda}
\def\la{\lambda}
\def\ck{\chi_k}

\markboth{D. L\'OPEZ NACIR and F.D. MAZZITELLI} 
{Renormalization for Quantum Fields with Modified Dispersion Relations}

%
\catchline{}{}{}{}{}
%

\title{ON THE RENORMALIZATION PROCEDURE FOR QUANTUM FIELDS WITH MODIFIED DISPERSION RELATIONS IN CURVED SPACETIMES}

\author{D. L\'OPEZ NACIR and F.D. MAZZITELLI}

\address{Departamento de Fisica J.J. Giambiagi, Facultad de Ciencias Exactas y Naturales,
UBA, \\ Ciudad Universitaria, Pabell\'on 1,
1428 Buenos Aires, Argentina.}

\maketitle

\begin{history}
\received{Day Month Year}
\revised{Day Month Year}
\end{history}

\begin{abstract}

We review our recent results on the renormalization procedure for a free quantum
scalar field with modified dispersion relations in curved
spacetimes.  For dispersion relations containing up to $2s$ powers of the spatial momentum,
the subtraction necessary to renormalize $\langle\phi^2\rangle$ and $\langle T_{\mu\nu}
\rangle$ depends on $s$.
We first describe our previous analysis for spatially flat
Friedman-Robertson-Walker and Bianchi type I metrics. Then we present
a new power counting analysis for general background metrics in the weak
field approximation.
\end{abstract}
\ccode{PACS numbers: 04.60.Bc, 04.62.+v, 11.10.Gh}
\subsection*{}\vspace{-0.5 cm}
It has been argued that trans-Planckian effects could be relevant
in the early universe and in the context of black hole physics. As
a phenomenological approach to investigate physics near the Planck
scale (or near a critical scale for which new physics could show
up), it is useful to analyze the consequences of assuming modified
dispersion relations (MDR) for the quantum fields,  in order to
assess the robustness of the predictions obtained in semiclassical
gravity. The MDR will of course affect the structure of the
quantum field theory, in particular its renormalizability. In the
semiclassical approximation, the renormalization of the stress
tensor is crucial to evaluate the backreaction of quantum fields.

The renormalization procedure for quantum fields satisfying the
standard dispersion relation in curved backgrounds is well
established.\cite{waldBiFu} Indeed, there are well known covariant
methods of renormalization that can be implemented in principle in
any spacetime metric. When applied to the expectation value of the
square of the field $\langle \phi^2\rangle $, or to the mean value
of the stress tensor $\langle T_{\mu\nu} \rangle$, one can obtain
the associated renormalized quantities by making the subtractions:
 \begin{subequations}\label{renor} \begin{align} \label{phiren}
\langle \phi^2 \rangle_{\mathrm{ren}}&=\langle \phi^2
\rangle-\langle \phi^2
\rangle^{(0)}...-\langle \phi^2
\rangle^{(2 i_{max})},\\
\langle T_{\mu\nu} \rangle_{\mathrm{ren}}&=\langle T_{\mu\nu}
\rangle-\langle T_{\mu\nu} \rangle^{(0)}...-\langle T_{\mu\nu}
\rangle^{(2 j_{max})},\label{tren}\end{align}\end{subequations}
where a superscript $2l$ denotes the terms of adiabatic order $2l$
of the corresponding expectation value (i.e., the terms containing
$2l$ derivatives of the metric).  For the usual dispersion
relation, it is well known that in $n$ dimensions the subtraction
involves up to $2i_{max}=2\, \mathrm{int}(n/2-1)$ for $\langle
\phi^2 \rangle$ and $2j_{max}=2\, \mathrm{int} (n/2)$ for the
stress tensor, where $\mathrm{int}(x)$ is the integer part of $x$.

The case of MDR can be consistently studied in the framework of the
Eintein-Aether theory.\cite{eatheory} In this theory, the general
covariance is preserved by introducing a dynamical vector field
$u^{\mu}$ called the aether field, which is constrained  to take a
non-zero timelike value, $u^{\mu}u_{\mu}=-1$. In the semiclassical
approximation, both the aether field and the spacetime metric are assumed to be
classical. The
action for a massive quantum scalar field $\phi$ can be written as\cite{lemoine}
 \be \label{accionphi}S_{\phi}=-\frac{1}{2}\int d^n x
\sqrt{-g}\lb \partial^{\mu}\phi\partial_{\mu}\phi+(m^2+\xi R)\phi^2+2\sum_{s,p\leq s} b_{sp}
(\mathcal{D}^{2s}\phi)(\mathcal{D}^{2p}\phi)\rb,\ee
where
$g=det(g_{\mu\nu})$, $R$ the Ricci scalar and 
$\mathcal{D}^{2}\phi\equiv\perp_{\mu}^{\lambda}\nabla_{\lambda}(\perp_{\gamma}^{\mu}\nabla^{\gamma}\phi)$ (with $\perp_{\mu\nu}\equiv g_{\mu\nu}+ u_{\mu} u_{\nu}$ and
$\nabla_{\mu}$  the derivative operator associated with
$g_{\mu\nu}$). The last term in Eq.~(\ref{accionphi}) gives rise to the MDR.

It has been realized that some non-trivial issues arise in the
renormalization procedure. On the one hand, the structure of the
counterterms could be different from the case of the standard dispersion relation.\cite{RW12}\cdash\cite{rinaldi}
Indeed, as the scalar field couples not only to the metric but
also to the aether field,  from a general effective field theory
perspective one can expect that new counterterms constructed with
both the metric and the aether field will be required. On the
other hand, the presence of higher spatial derivatives affects the
singularity structure of the propagator, and one is led to the
question of whether higher values of $s$ in Eq.~(\ref{accionphi})
imply milder divergences in the unrenormalized quantities or not.
In other words, given a MDR, we are interested in knowing up to
which adiabatic order the subtractions in Eq.~(\ref{renor}) have to be
carried out to get finite, physically meaningful expectation
values. In the case of interacting quantum fields in Minkowski
spacetime, it has been shown that higher spatial derivatives
improve the UV behavior of Feynman diagrams.\cite{Anselmi} Here,
we will show that while such improvement also occurs for $\langle
\phi^2 \rangle$, the opposite holds for $\langle
T_{\mu\nu}\rangle$.

For scalar fields propagating in a spatially flat
Friedman-Robertson-Walker (FRW) spacetime of $n$ dimensions, the
extension of the adiabatic subtraction scheme based in a WKB
expansion of the field modes has been considered in
Ref.~\refcite{RW12}. The Fourier modes of the scaled field
$\chi=C^{(n-2)/4}(\eta)\phi$ satisfy \be
\chi_k''+\lb(\xi-\xi_n)RC(\eta)+\om_k^2\rb\chi_k=0, \label{PXXP}\ee
where $\sqrt{C(\eta)}$ is the scale factor, primes stand for derivatives
with respect to the conformal time $\eta$, $\xi_n=(n-2)/(4n-4)$,
and \be \om^2_k= |\vec{k}|^2+C(\eta)\lb m^2+2\sum_{s,p\leq
s}(-1)^{s+p}\,b_{sp}\,
\lp\frac{|\vec{k}|}{\sqrt{C(\eta)}}\rp^{2(s+p)}\rb. \label{dis} \ee To
get the WKB expansion, we express $\chi_k$ as \be \ck=
\frac{1}{\sqrt{ 2 W_k}}\exp\lp -i\int^\eta
W_k(\tilde\eta)d\tilde\eta\rp ,\label{chi} \ee and substitute this
into Eq.~(\ref{PXXP}) to obtain a nonlinear differential equation
for $W_k^2$. Solving this equation iteratively, it can be shown that the $2l-$adiabatic order
of $W_k^2$ scales as $\om_k^{2-2l}$. After substituting
Eq.~(\ref{chi}) into $\langle \phi^2\rangle$ and $\langle
T_{\mu\nu}\rangle$, one can determine whether  a given adiabatic
order of these expectation values is finite or not. In this way,
for a MDR such that the frequency behaves as $\om\sim |\vec{k}|^s$
for large values of $|\vec{k}|$, one can show that
divergences appear up to \be\label{jmaxhomog}  2i_{max}=2\,
\mathrm{int}\left(\frac{n-1}{2 s}-\frac{1}{2}\right),\,\,\,
2j_{max}=2\, \mathrm{int}\left(\frac{1}{2}+\frac{n-1}{2
s}\right).\ee In Ref.~\refcite{RW12} the WKB expansion of the
stress tensor was computed up to the fourth adiabatic order for
the class of MDR given in Eq.~(\ref{dis}).\footnote{As
Eq.~(\ref{jmaxhomog}) indicates, for $n=4$ the fourth adiabatic
order is convergent when $s\geq 2$, and the second order is
convergent when $s\geq 4$. However, there are subtle points in the
renormalization procedure related to the trace anomaly.} It was
shown that these adiabatic orders can be absorbed into a
redefinition of the gravitational bare constants of the theory, as
for the usual dispersion relation (i.e., only geometric counterterms are needed). However, this simple result is
due to the symmetries of the spatially flat FRW metric.

In Bianchi type I spacetimes, the WKB expansion can be obtained in
a completely analogous way, and Eq.~(\ref{jmaxhomog}) also applies
in this case.\cite{Bianchi} However, for these anisotropic
metrics, one can show that new counterterms are necessary, which
involve the timelike vector field in addition to the metric.\cite{Bianchi} For instance, a term proportional to
$(\nabla_{\mu}u^{\mu})^2$ in the aether Lagrangian is
needed to absorb the divergences in $\langle
T_{\mu\nu}\rangle^{(2)}$ (in addition to the usual Einstein-Hilbert
action).  The point is that in a spatially flat FRW background these new
counterterms are indistinguishable from the usual ones.
Concretely, once evaluated in this background, the stress tensor
obtained from the variation of the most general action for the
aether field containing two derivatives, turns out to be
proportional to the Einstein tensor.

Currently, there are strong constraints on the parameters associated to terms containing
two derivatives of the aether field.\cite{Jacobson} Therefore, the new counterterms of second adiabatic
order should be carefully chosen to make the theory consistent with observation.\cite{Bianchi}

The values of $2i_{max}$ and $2j_{max}$ in Eq.~(\ref{jmaxhomog})
are a peculiarity of the spatially homogeneous backgrounds
considered so far. To see this, let us consider a general
background in the weak field approximation, $
g_{\mu\nu}=\eta_{\mu\nu}+h_{\mu\nu}\,,\,u_{\mu}=\delta_{\mu}^0+v_{\mu}.$
By keeping only linear terms in $h_{\mu\nu}$ and $v_{\mu}$, an
integral expression of the Feynman propagator $G_{F}(x,x')$ for
the scalar field can be obtained perturbatively:
$G_F=G_F^0+G_F^1$,  where the superscripts refer to the order in
$h_{\mu\nu}$ and $v_{\mu}$,
\begin{subequations}\begin{align}
G^{0}_F(x,x')&=(2\pi)^{-n}\int d^n k e^{i k(x-x')}[-{k_0}^2+\om^2(|\vec{k}|^2)]^{-1},\\
G^1_F(x,x')&=-\int d^n y
G_F^{0}(x,y)\mathcal{F}(y)G^0_F(y,x').
\end{align}\end{subequations} Here $\mathcal{F}$ is  an  operator linear in the perturbation fields, and
$ \om^2(|\vec{k}|^2)=m^2-i\epsilon+|\vec{k}|^2+2\sum_{s,p\leq s}b_{sp} (-1)^{s+p}\ |\vec{k}|^{2(s+p)}$.

The expectation value $\langle \phi^2 \rangle$ is given by the
coincidence limit of $\mathrm{Im} G_F$.  Analogously,  $\langle
T_{\mu\nu} \rangle$ can be expressed as the coincidence limit of a
derivative operator applied to $\mathrm{Im} G_{F}$. In this way, one can obtain integral
expressions for  both $\langle \phi^2 \rangle$ and  $\langle
T_{\mu\nu}\rangle$. Using an expansion in derivatives of the perturbation fields, one can
study up to which adiabatic order these quantities contain
divergences. For a MDR such that $\om\sim |\vec{k}|^s$ for large
values of $|\vec{k}|$,  a power counting analysis yields\cite{Campodebil} \be\label{jmaxCampodebil} 2i_{max}=2\,
\mathrm{int}\left(\frac{n-1-s}{2}\right),\,\,\, 2j_{max}=2\,
\mathrm{int}\left(\frac{n-1+s}{2}\right).\ee  The value of $2i_{max}$ is
now generally larger than the one given in Eq.~(\ref{jmaxhomog}),
although it also {\it decreases} with $s$. However, contrary to the
previous case, $2j_{max}$ {\it increases} with $s$. Therefore we
conclude that, in the weak field approximation, for a general background the subtraction in
Eq.~(\ref{renor}) should be performed up the adiabatic orders
$2i_{max}$  and $2j_{max}$  given in Eq.~(\ref{jmaxCampodebil}).
In particular, in order to renormalize the semiclassical Einstein-Aether equations, it will be necessary to introduce all possible counterterms constructed with $g_{\mu\nu}$ and $u_{\mu}$, up to the $2j_{\max}-$adiabatic order. It would be interesting to check if these results remain valid beyond the weak field approximation.

\section*{Acknowledgments}
This work has been supported by  Universidad de Buenos Aires,
CONICET and ANPCyT.



\begin{thebibliography}{0}    


\bibitem{waldBiFu}  N. D. Birrell and P. C. W. Davies, {\it Quantum Fields in
Curved Space} (Cambridge University Press, Cambridge, 1982).

\bibitem{eatheory} T. Jacobson and D. Mattingly, {\it Phys. Rev. D }{\bf 63},
041502(R) (2001); ibidem {\it D} {\bf 64}, 024028 (2001).

\bibitem{lemoine} M. Lemoine, M. Lubo, J. Martin and J. P. Uzan, {\it Phys. Rev. D } {\bf 65},
023510 (2001).

\bibitem{RW12} D. L\'opez Nacir, F. D. Mazzitelli, and C. Simeone, {\it Phys. Rev. D } {\bf 72},
124013 (2005); D. L\'opez Nacir and F. D. Mazzitelli, {\it Phys. Rev. D} {\bf 76}, 024013 (2007).

\bibitem{Bianchi} D. L\'opez Nacir and F. D. Mazzitelli, {\it Phys. Rev. D } {\bf 78},
044001 (2008).

\bibitem{rinaldi} M. Rinaldi, {\it Phys. Rev. D } {\bf 76}, 104027 (2007); ibidem {\it D} {\bf 78},024025 (2008).

\bibitem{Anselmi} D. Anselmi and M. Halat {\it Phys. Rev. D } {\bf 76}, 125011 (2007).

\bibitem{Jacobson}  T. Jacobson,  {\it arXiv:0801.1547 [gr-qc]}.

\bibitem{Campodebil}  D. L\'opez Nacir and F. D. Mazzitelli, in preparation.


\end{thebibliography}
\end{document}